\documentclass[aps,twocolumn,nofootinbib,superscriptaddress,preprintnumbers]{revtex4}
\usepackage{slashed}
\usepackage{graphicx}
\usepackage{times}
\usepackage{epsfig}
\usepackage{amsmath}
\usepackage{amsfonts}
\usepackage{amssymb}
\usepackage{url}
\usepackage{subfigure}
\newcommand{\be}{\begin{equation}}
\newcommand{\ee}{\end{equation}}
\newcommand{\bea}{\begin{eqnarray}}
\newcommand{\eea}{\end{eqnarray}}

\begin{document}
\preprint{KEK-TH-1198}
\pagestyle{plain}
\title{Unparticle Dark Matter}
\author{Tatsuru Kikuchi}
\email{tatsuru@post.kek.jp}
\affiliation{Theory Division, KEK,
1-1 Oho, Tsukuba, 305-0801, Japan.}
\author{Nobuchika Okada}
\email{okadan@post.kek.jp}
\affiliation{Theory Division, KEK,
1-1 Oho, Tsukuba, 305-0801, Japan.}
\affiliation{Department of Particle and Nuclear Physics,
The Graduate University for Advanced Studies,\\
Oho 1-1, Tsukuba, Ibaraki, 305-0801, Japan}
\begin{abstract} 
Once a parity is introduced in unparticle physics, 
 under which unparticle provided in a hidden conformal sector 
 is odd while all Standard Model particles are even, 
 unparticle can be a suitable candidate for the cold dark matter (CDM) 
 in the present universe through its coupling 
 to the Standard Model Higgs doublet. 
We find that for Higgs boson mass in the range, 
 114.4 GeV $\lesssim  m_h \lesssim$ 250 GeV, 
 the relic abundance of unparticle with mass 
 50 GeV $\lesssim  m_{\cal U}  \lesssim$ 80 GeV 
 can be consistent with the currently observed CDM density. 
In this scenario, Higgs boson with mass $m_h \lesssim 160$ GeV 
 dominantly decays into a pair of unparticles and 
 such an invisible Higgs boson may be discovered 
 in future collider experiments. 
\end{abstract}
\maketitle
Existence of the dark matter (DM) is now strongly supported 
 by various observations of the present universe,  
 in particular, the Wilkinson Microwave Anisotropy Probe (WMAP) 
 satellite \cite{WMAP} have determined the various cosmological 
 parameters with greater accuracy. 
The relic abundance of cold dark matter (CDM) 
 is estimated to be (in 2$\sigma$ range) 
\begin{eqnarray}
0.096 \le \Omega_{\rm CDM} h^2 \le 0.122 \; . 
 \label{WMAP} 
\end{eqnarray}
To clarify the identity of a particle as cold dark matter is 
 still a prime open problem both in particle theory and cosmology.

Absence of any suitable candidate of cold dark matter 
 in the Standard Model (SM) suggests the existence 
 of new physics beyond the SM in which a dark matter candidate  is implemented. 
The most promising candidate of CDM is the so-called weakly 
 interacting massive particle (WIMP). 
Once the stability of WIMP is ensured by some symmetry (parity), 
 its relic abundance can naturally be consistent with the WMAP data 
 for WIMP mass and its typical interaction scales 
 around the electroweak scale.  
This scale is accessible to future collider experiments 
 such as  the Large Hadron Collider (LHC) at CERN 
 which will be in store for its operation next year. 
A large missing energy associated with WIMP DM production 
 is one of the important keys to discover new physics 
 at collider experiments.  
There have been proposed the WIMP DM candidates 
 in several new physics models, 
 such as neutralino as the lightest sparticle 
 in supersymmetric model with R-parity, 
 the neutral heavy vector boson 
 in the littlest Higgs model with T-parity \cite{LHwT}, 
 the lightest Kaluza-Klein particle 
 in the universal extra dimension model \cite{LKP}
 and so on.

In this letter, we propose a new candidate for CDM 
 in the context of a new physics model recently proposed 
 by Georgi \cite{Georgi:2007ek}, "unparticle". 
We introduce a ${\mathbb Z}_2$ parity under which unparticle 
 is odd while all Standard Model particles are even. 
The unparticle, which is provided by a hidden conformal sector 
 and is originally massless, obtains masses associated with 
 the electroweak symmetry breaking through its coupling to 
 the SM Higgs doublet. 
We find that the unparticle can be a suitable candidate for CDM 
 through the coupling. 
In addition, in our scenario 
 the SM Higgs boson can invisibly decay into 
 a pair  of unparticles with a large branching ratio. 

Unparticle provided in a hidden conformal sector 
 could posses strange properties, especially 
 in its energy distributions. 
A concrete example which can proved unparticle 
 was discussed by Banks-Zaks \cite{Banks:1981nn} (BZ) many years ago, 
 where introducing a suitable number of massless fermions, 
 theory reaches a non-trivial infrared fixed point and 
 a conformal theory can be realized at low energy 
\footnote{Our present analysis does not depend on the model behind unparticle.
We suppose a more general theory than the BZ theory for the model behind the unparticle, 
where the unparticle provided as a composite state in low energy effective theory, 
like baryons in QCD.
In such a theory, we may expect  that a low energy effective theory includes a global symmetry
like the baryon number in QCD and a composite state has a non-trivial charge under it
like the baryon number of proton and neutron.
We assume such situation for the unparticle and introduce a ${\mathbb Z}_2$ symmetry 
under which the unparticle is odd.
}.
After the Georgi's proposal, it has been paid a lot of interests 
 in the unparticle physics and various studies on 
 the unparticle physics in scope of the LHC, cosmology, etc. 
 have been developed in the literature. 
%

Now we begin with a very brief review of the basic structure 
 of the unparticle physics. 
First, we introduce a coupling between the new physics operator 
 ($\cal{O}_{\rm UV}$) with dimension $d_{\rm UV}$ 
 and the Standard Model one (${\cal O}_{\rm SM}$) with dimension $n$, 
\bea
 {\cal L} = \frac{c_n}{M^{d_{\rm UV}+n-4}} 
     \cal{O}_{\rm UV} {\cal O}_{\rm SM} ,  
\eea
where $c_n$ is a dimension-less constant, and $M$ is the energy scale 
 characterizing the new physics. 
This new physics sector is assumed to become conformal 
 at a scale $\Lambda_{\cal U}$, and 
 the operator $\cal{O}_{\rm UV}$ flows to the unparticle operator 
 ${\cal U}$ with dimension $d_{\cal U}$. 
In low energy effective theory, we have the operator of the form 
 (here we consider scalar unparticle, for simplicity), 
\bea
{\cal L}=c_n 
 \frac{\Lambda_{\cal U}^{d_{\rm UV} - d_{\cal U}}}{M^{d_{\rm UV}+n-4}}   
 {\cal U} {\cal O}_{\rm SM} 
\equiv 
  \frac{1}{\Lambda^{d_{\cal U}+ n -4}}  {\cal U} {\cal O}_{\rm SM},  
\label{basic-int}
\eea 
where the scaling dimension of the unparticle ($d_{\cal U}$) have been 
 matched by $\Lambda_{\cal U}$ which is induced the dimensional 
 transmutation, and $\Lambda$ is the (effective) cutoff scale 
 of low energy effective theory. 
Interestingly, $d_{\cal U}$ is not necessarily to be integer, 
 but can be any real number or even complex number. 
In this paper we consider the scaling dimension in the range, 
 $1 \leq  d_{\cal U} <  2$, for simplicity. 
It was found in Ref.~\cite{Georgi:2007ek} that, 
 by exploiting scale invariance of the unparticle, 
 the phase space for an unparticle operator with the scale dimension 
 $d_{\cal U}$ and momentum $p$ is the same as the phase space 
 for $d_{\cal U}$ invisible massless particles, 
\begin{eqnarray}
d \Phi_{\cal U}(p) = 
 A_{d_{\cal U}} \theta(p^0) \theta(p^2)(p^2)^{d_{\cal U}-2} 
 \frac{d^4p}{(2\pi)^4} \,,
\label{Phi}
\end{eqnarray}
where
\begin{eqnarray}
A_{d_{\cal U}} = \frac{16 \pi^{\frac{5}{2}}}{(2\pi)^{2 d_{\cal U}}}
\frac{\Gamma(d_{\cal U}+\frac{1}{2})}{\Gamma(d_{\cal U}-1) 
\Gamma(2 d_{\cal U})}.
\label{A}
\end{eqnarray}
%
Also, based on the argument on the scale invariance, 
 the (scalar) propagator for the unparticle was suggested to be 
 \cite{Georgi:2007si, Cheung:2007ue} 
\begin{eqnarray}
 \frac{A_{d_{\cal U}}}{2\sin(\pi d_{\cal U})}
 \frac{i}{(p^2)^{2-d_{\cal U}}} 
 e^{-i (d_{\cal U}-2) \pi}  . 
\label{propagator}
\end{eqnarray}
%
Because of its unusual mass dimension, 
 unparticle wave function behaves as 
 $\sim  (p^2)^{(d_{\cal U}-1)/2}$ 
 (in the case of scalar unparticle).

Now let us impose a ${\mathbb Z}_2$ parity 
 under which unparticle is odd while all SM particles are even, 
 so that unparticle should appear in a pair in interaction terms. 
Among many possibilities, we focus on the interaction term 
 between unparticles and 
 the Standard Model Higgs doublet ($H$) such as 
\bea 
 {\cal L}_{\rm int}  = - \frac{\lambda}{\Lambda^{2d_{\cal U} - 2}}  
 {\cal U}^2 \left( H^\dagger H \right), 
\label{U-Higgs-Org}
\eea
where $\lambda$ is a real and positive dimensionless coefficient. 
Note that this is the lowest dimensional operator 
 among all possible operators between a pair of unparticles 
 and the SM particles. 
Thus, this operator would be the most important one 
 in unparticle phenomenology at low energies, 
 at least, it is so in our discussion on unparticle dark matter 
 for $\Lambda \gtrsim 1$ TeV, for example. 
Although unparticle is originally provided by a hidden conformal sector 
 and is massless, it obtains mass through this interaction 
 once the Higgs doublet develops the vacuum expectation value (VEV), 
 $\left< H  \right> = v/\sqrt{2}$ ($v=$ 246 GeV), 
 breaking the electroweak symmetry. 
After the symmetry breaking, we have 
\bea 
 {\cal L}_{\rm int}  = 
  - \frac{1}{2} m_{\cal U}^{4-2 d_{\cal U}} {\cal U}^2
   \left( 
   1+ 2 \frac{h}{v}+ \frac{h^2}{v^2} 
   \right) , 
\label{U-HiggsInt}
\eea
where 
$m_{\cal U}=(\sqrt{\lambda} v/\Lambda^{d_{\cal U}-1})^{1/(2-d_{\cal U})}$ 
 is the unparticle mass,
 and $h$ is the physical Standard Model Higgs boson. 
For $d_{\cal U} \sim 1$ unparticle has mass around the electroweak scale 
 and interactions with Higgs boson characterized also 
 by the electroweak scale.
The parity we have introduced ensures the stability of unparticle. 
These are ideal situations for unparticle 
 to be the WIMP dark matter.

Our scenario shares similar structures with some simple modes 
 for dark matter \cite{NSM}, 
 where the gauge singlet scalar is introduced into the SM 
 and can be a suitable candidate for dark mater 
 through couplings to Higgs boson. 
The crucial difference of unparticle from such a singlet scalar 
 is that unparticle is originally massless 
 because of the conformal invariance of a hidden sector. 
The absence of mass term reduces the number of free parameters 
 involved in dark matter physics and as a result, 
 we can analyze the relic density of unparticle dark matter 
 as a function of only unparticle mass ($m_{\cal U}$) and 
 Higgs boson mass ($m_h$), as we will see later. 

Now let us evaluate the relic density of unparticle dark matter. 
In our analysis, we consider the case $d_{\cal U} \sim 1$, 
 for simplicity, where unparticle is almost identical 
 to a gauge singlet scalar
 %
We can expect that even for a general $d_{\cal U}$ in the range, $1 \leq d_{\cal U} < 2$, 
our results will remain almost the same in the following reasons. 
First, the phase space factor $A_{d_{\cal U}}$ is 
 a slowly varying function of $d_{\cal U}$. 
Second, the unparticle dark matter decouples from thermal bath 
 in non-relativistic regime, where the most important  factor 
 to fix the decoupling temperature is the Boltzmann factor 
 $e^{-m_{\cal U}/T}$ independent of $d_{\cal U}$. 
 \footnote{
In precise, the thermal history of the unparticle is still an issue under discussion.
In this paper, since we consider a massive unparticle and its decoupling nature in
non-relativistic regime, we implicitly assume that the thermal distribution of unparticle
is almost the same as usual WIMP dark matter with the Boltzmann suppression factor.
One the other hand, in relativistic regime, it has been demonstrated in \cite{thermal-un}.
that thermal distribution of the unparticle is quite different from usual relativistic particle.
It is an interesting issue to find a correct formula which smoothly connects relativistic
regime with non-relativistic one (that we expect). We leave this issue for future study.
}
Moreover, in non-relativistic regime, the unparticle wave function 
 behaves as $m_{\cal U}^{d_{\cal U}-1}$ and 
 the interaction terms in Eq.~(\ref{U-HiggsInt}) 
 becomes independent of $d_{\cal U}$ in momentum space. 

The relic abundance of the dark matter is obtained by solving 
 the following Boltzmann equation \cite{Kolb:1990vq},
\begin{eqnarray}
 \frac{d Y}{d x}
 =
 -\frac{\langle \sigma {\rm v} \rangle}{H x}s
 \left( Y^2 - Y_{\rm eq}^2 \right)~,
 \label{eq:Boltzmann}
\end{eqnarray}
 where $Y = n/s$ is the yield of the dark matter defined by the ratio of
 the dark matter density ($n$) to the entropy density of the universe 
 ($s = 0.439g_*m_{\cal U}^3/x^3$), $g_* = 86.25$, and 
 $x \equiv m_{\cal U}/T$ ($T$ is the temperature of the universe). 
The Hubble parameter is given by 
 $H = 1.66g_*^{1/2} m_{\cal U}^2 m_{\rm Pl}/x^2$, 
 where $m_{\rm Pl} = 1.22\times 10^{19}$ GeV is the Planck mass, 
 and the yield in the equilibrium $Y_{\rm eq}$ is written 
 as $Y_{\rm eq} = (0.434/{g_*}) x^{3/2} e^{-x}$.
%
%
After solving the Boltzmann equation with the thermal averaged 
 annihilation cross section $\langle\sigma {\rm v}\rangle$, 
 we obtain the present abundance of dark matter ($Y_\infty$). 
With a good accuracy, the solution of Eq.~(\ref{eq:Boltzmann}) 
 is approximately given as \cite{Kolb:1990vq} 
\begin{eqnarray}
 \Omega h^2
 &=& 
 \frac{1.07\times 10^9 x_f{\rm GeV}^{-1}}
 {\sqrt{g_*}m_{\rm PL}\langle\sigma {\rm v}\rangle} ,
\end{eqnarray}
 where $x_f = m_{\cal U}/T_f$ is the freeze-out temperature 
 for the dark matter and given as $x_f = \ln(X) - 0.5\ln(\ln(X))$ 
 with $X = 0.038\cdot (1/g_*^{1/2})m_{\rm PL}m_{\cal U}\langle 
 \sigma {\rm v}\rangle$. 

The unparticle dark matter annihilates into the SM particles 
 through its interaction to Higgs boson in Eq.~(\ref{U-HiggsInt}). 
Since this annihilation occurs in the s-wave, 
 the thermal averaged annihilation cross section 
 $\langle\sigma  v \rangle$ is simply given by  
\bea
 \langle\sigma  v \rangle = 
 \sum_{IJ} \left. \sigma   v \right|_{IJ} ,
\eea 
where $I, J$ stand for SM particles 
 in each possible annihilation process 
 ${\cal U}{\cal U} \to IJ$. 
When $m_{\cal U} \leq m_h$, possible annihilation processes 
 of unparticle dark matters are ${\cal U} {\cal U} \to h \to IJ$, 
 where $IJ = f \bar{f}$, $W^+ W^-$, $ZZ$, etc. 
In our analysis, off-shell states for $IJ=W^+ W^-$ and $ZZ$ 
 are also taken into account. 
When $m_{\cal U} > m_h$, the process 
 $IJ= h h$ should be added into the annihilation processes. 
However, we find that the WMAP allowed region 
 appears most for $m_{\cal U} < m_h$, 
 and it is sufficient to consider only the processes 
 mediated by Higgs boson in the s-channel. 
In this case, the annihilation cross section 
 can be simply described as  
\bea
\left. \sigma  v \right|_{IJ} 
 = 4 \frac{m_{\cal U}^3}{v^2} 
  \frac{ \Gamma(h \to IJ)\Big|_{m_h =2 m_{\cal U}} }
 {(4 m_{\cal U}^2 - m_h^2)^2 + m_h^2 \Gamma_h^2} ,
\label{crosssec}
\eea
where $\Gamma(h \to IJ)$ is the SM Higgs boson partial decay width 
 into $IJ$, and the subscript $m_h =2 m_{\cal U}$ means 
 to replace $m_h$ into $2 m_{\cal U}$ 
 in the formula of the Higgs boson partial decay width. 
When $m_h > 2 m_{\cal U}$, the partial decay width, 
\bea
\Gamma(h \to {\cal U} {\cal U}) =
  \frac{1}{8 \pi m_h} \frac{m_{\cal U}^4}{v^2} 
  \sqrt{1- 4 \frac{m_{\cal U}^2}{m_h^2}} ,
\eea
 should be added in the Higgs boson total decay width $\Gamma_h$.  
Note that the relic density of the unparticle dark matter 
 can be determined by only two free parameters, $m_{\cal U}$ and $m_h$. 

\begin{figure}[htbp]
\begin{center}
{\includegraphics[width=0.8\columnwidth]{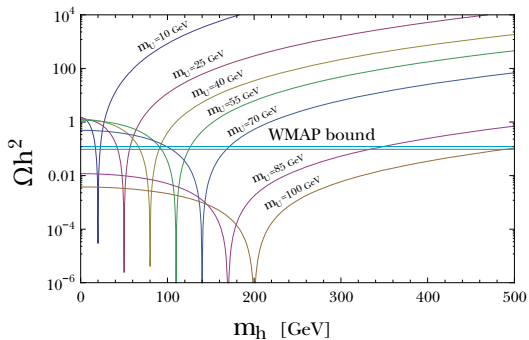}}
\caption{
The relic abundance of the unparticle dark matter 
 as a function of the Higgs boson mass 
 for fixed unparticle masses, together with the WMAP measurements, 
 $0.096 \leq \Omega_{\rm CDM} h^2 \leq 0.122$. 
Each curve corresponds to the unparticle mass, 
 $m_{\cal U} =10,~25,~40,~55,~70,~85,~100$ GeV. 
}
\label{Fig1}
\end{center}
\end{figure}
\begin{figure}[htbp]
\begin{center}
{\includegraphics[width=0.8\columnwidth]{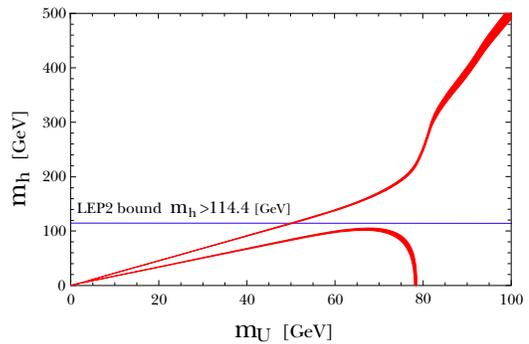}
\caption{
The contour plot of the relic abundance of the unparticle dark matter 
 $\Omega h^2$ in ($m_{\cal U}$, $m_h$)-plane. 
The shaded thin area is the allowed region 
 for the WMAP measurements, 
 $0.096 \le \Omega_{\cal U} h^2 \le 0.122$, 
 at $2\sigma$ confidence level. 
}
\label{Fig2}}
\end{center}
\end{figure}

\begin{figure}[htbp]
\begin{center}
{\includegraphics[width=0.8\columnwidth]{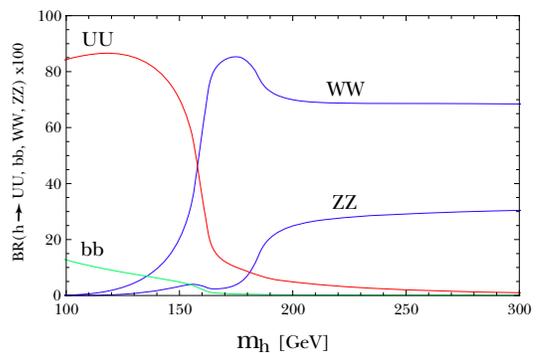}
\caption{
Branching ratios of the Higgs boson decays 
 as a function of the Higgs mass, 
 along the WMAP allowed region for $m_h > 2 m_{\cal U}$ 
 in Fig.~\ref{Fig2}.
} 
\label{Fig3}
}
\end{center}
\end{figure}

In Fig.~\ref{Fig1}, the relic abundance of the unparticle dark matter 
 are depicted as a function of the Higgs boson mass 
 for fixed unparticle masses, together with the WMAP result. 
The relic abundance sharply falls down at Higgs boson pole,  
 $m_h=2 m_{\cal U}$, because of the resonance, 
 as can be easily understood from Eq.~(\ref{crosssec}). 
Therefore, the WMAP consistent region appears 
 in both sides of the Higgs pole. 
The cross section becomes larger as $m_{\cal U}$ is raised for fixed $m_h$, 
 so that the WMAP allowed region for $m_h < 2 m_{\cal U}$ 
 eventually disappears. 
This growth of the annihilation cross section is 
 related to the unitarity violation, 
 since the original interaction in Eq.~(\ref{U-Higgs-Org}) 
 is higher dimensional and the cross section becomes larger 
 as energy, in other word, $m_{\cal U}$ becomes large. 
For $d_{\cal U}=1$, this corresponds to rasing a coupling $\lambda$. 
The allowed region always exists for $m_h > 2 m_{\cal U}$, 
 because the annihilation cross section is suppressed 
 for a large $m_h$.

The WMAP allowed region on ($m_{\cal U}$, $m_h$)-plane is shown 
 in Fig.~\ref{Fig2}. 
The lower bound on the Higgs boson mass by LEP2 \cite{LEP2} 
 excludes the WMAP allowed region for $m_h < m_{\cal U}$ 
 and the region $m_{\cal U} \lesssim 50$ GeV. 
From $m_{\cal U} \simeq$ 80 GeV, 
 the Higgs boson mass starts growing quickly 
 since the annihilation process into a real $W$-boson pair 
 in the final state opens up and the annihilation cross section 
 becomes large from the threshold. 
Light Higgs boson mass $m_h \lesssim$ 250 GeV 
 is favored from the electroweak precision measurements \cite{PDG}, 
 so that the unparticle mass is constrained to be in the range, 
 50 GeV $\lesssim m_{\cal U} \lesssim$ 80 GeV. 

As shown in Fig.~\ref{Fig2}, the region consistent with 
 both the WMAP data and the Higgs boson mass bound by LEP2 
 appears only for $m_h > 2 m_{\cal U}$, so that 
 Higgs boson can decay into a pair of unparticles. 
The branching ratio is depicted in Fig.~\ref{Fig3}. 
In fact, for $m_h \lesssim 160$ GeV, Higgs boson dominantly 
 decays into unparticle dark matters. 
Even for $m_h=200$ GeV, the branching ratio 
 of invisible Higgs boson decay is sizable, 
 ${\rm BR}(h \to {\cal U}{\cal U}) \simeq 8.5$ \%. 
Besides our scenario and simply extended SM models \cite{NSM}, 
 the invisible Higgs boson decay has been discussed 
 in Majoron models \cite{Majoron}, extra-dimension models \cite{radion} 
 and the little Higgs model with T-parity \cite{LHinv}. 
When Higgs boson dominantly  decays into the invisible mode, 
 the Higgs boson search at LHC would be more challenging. 
However, there are several ideas to search the invisibly decaying 
 Higgs boson through its associated productions 
 with weak bosons \cite{associatedW} or top quarks \cite{associatedtop} 
 and its production through weak boson fusion \cite{WBF}. 
On the other hand, at International Linear Collider (ILC), 
 the search for such an invisible Higgs boson 
 and the measurement of its invisible decay width are easier 
 through the final state fermions from recoiled Z-boson decay. 

In summary, we have investigated the possibility of 
 unparticle dark matter. 
Imposing the ${\mathbb Z}_2$ parity for unparticle 
 and hence ensuring the stability of unparticle, 
 we have introduced the coupling between 
 unparticles and the SM Higgs doublets. 
Associating with the electroweak symmetry breaking, unparticle 
 obtains mass and becomes the WIMP dark matter candidate. 
We have evaluated the relic abundance of unparticle dark matter 
 and found the WMAP allowed region with the unparticle mass 
 around the electroweak scale. 
Interestingly, in this allowed region, 
 Higgs boson can decay into a pair  of unparticle dark matters 
 with a sizable branching ratio, even this invisible decay mode 
 can be dominant. 
Such an invisible Higgs boson may be observed 
 in future collider experiments.  
It would be worth investigating indirect detections 
 of unparticle dark matter through cosmic rays originating 
 from unparticle pair annihilation 
 in the halo associated with our galaxy. 
Since this annihilation occurs in the s-wave, 
 the annihilation cross section does not suffer from 
 the suppression by low relative velocity 
 of colliding unparticles, as a result, 
 we can expect a sizable cosmic ray flux. 
Cosmic positron flux has been analyzed in Ref. \cite{Asano} 
 for the dark matter in the littlest Higgs model with T-parity. 
The annihilation processes of unparticles we have considered 
 are basically the same as those in the paper, 
 and we can apply the same arguments for the cosmic ray 
 from unparticle annihilation in the halo. 
In this letter, we have assumed scalar unparticle, for simplicity. 
It is easy to consider fermionic or vector unparticle 
 as the dark matter.
We will arrive at the same conclusions except different numerical factors 
 related with the representations under Lorentz group. 

\begin{center}
{\bf Acknowledgments}
\end{center}
The work of T.K. was supported by the Research Fellowship 
 of the Japan Society for the Promotion of Science (\#1911329). 
The work of N.O. is supported in part by 
 the Grant-in-Aid for Scientific Research from the Ministry 
 of Education, Science and Culture of Japan (\#18740170). 
%

\end{document}